%% file: main.tex
\documentclass[12pt,a4paper]{article}

\usepackage[utf8]{inputenc}
\usepackage[T1]{fontenc}
\usepackage[english]{babel}

\usepackage[margin=1in]{geometry}
\usepackage{setspace}
\usepackage{microtype}
\usepackage{parskip}

\usepackage{booktabs}
\usepackage{array}
\usepackage{tabularx}
\usepackage{graphicx}
\graphicspath{{figures/}}
\usepackage{caption}
\usepackage{amsmath,amssymb}
\usepackage{siunitx}
\usepackage[numbers,sort&compress]{natbib}
\usepackage[hidelinks]{hyperref}
\usepackage{url}

\usepackage{xcolor}

\title{Limited Marginal Benefit of Reasoning-Heavy LLM Deployment
  in ESG Narrative Scoring:
  A 4-Model Consensus Study on Japanese Listed Firms}
\author{Hiroyuki Kokubu\\\small Kansai University}
\date{\today}

\begin{document}
\maketitle

\begin{abstract}
\input{sections/00_abstract}
\end{abstract}


\input{sections/01_introduction}
\input{sections/02_methods}
\input{sections/03_results}
\input{sections/04_cost_analysis}
\input{sections/05_discussion}
\input{sections/06_limitations}
\input{sections/07_conclusion}

\bibliographystyle{plainnat}
\bibliography{bib/references}

\end{document}

%% file: sections/00_abstract.tex
Automated scoring of ESG narrative disclosures with large language models
(LLMs) is gaining traction, yet whether reasoning-heavy frontier models add
value commensurate with their cost remains empirically unsettled. We evaluate
this question on a corpus of ten Japanese listed firms across three rubric
axes --- quantitative targets, progress-tracking infrastructure, and
external-standard alignment --- using a four-model consensus design that
combines a reasoning-on frontier model with three reasoning-off
contemporaries. Across 120 firm$\times$axis$\times$model scores, the pooled
mean absolute deviation between the reasoning-on model and each reasoning-off
counterpart is 0.38 on a 5-point scale; only 2\% of pairwise comparisons
reach a two-point deviation, and none exceeds two points. Per-firm cost
accounting shows the reasoning-on arm alone costs roughly 5.6$\times$ as
much as the three-provider reasoning-off ensemble, for outcomes that differ
only within small margins. We conclude that in span-based ESG narrative
scoring, reasoning-heavy deployment does not materially improve outcomes
relative to reasoning-off consensus, while substantially increasing
operational cost. We discuss implications for cost-effective ESG auto-scoring
pipelines and LLM deployment governance in applied accountability settings.
An earlier version of this work is available on SSRN (Abstract ID 6683303).

%% file: sections/01_introduction.tex
\section{Introduction}
\label{sec:introduction}

\subsection{Automated ESG narrative scoring}
\label{sec:intro-esg-llm}

Sustainability disclosures published by listed firms have become
increasingly voluminous and heterogeneous, comprising integrated
annual reports, climate-related supplementary documents under the
Task Force on Climate-related Financial Disclosures (TCFD)
framework~\citep{tcfd2017}, and standalone sustainability reports.
Manual coding of these narratives at portfolio scale is costly and
slow, motivating a growing literature on automated AI-based ESG
evaluation~\citep{nakao2023aiesg} and, more recently, on
LLM-specific extraction and scoring approaches~\citep{zou2025esgreveal,ding2025eulesg},
with further extensions integrating retrieval-augmented generation
and external benchmark scores~\citep{yang2026esglens}. Recent work has
also examined zero-shot and few-shot LLM coding against human
benchmarks~\citep{wu2025aiannotator}, finding meaningful but
imperfect alignment with expert annotations and persistent
hallucination risk. Operational questions—\emph{which model to
deploy at what cost}—have received comparatively little empirical
attention.

\subsection{The reasoning-tier question}
\label{sec:intro-reasoning}

A new generation of ``reasoning'' models exposes an explicit
computation budget for chain-of-thought tokens that are billed in
addition to standard input and output tokens. Prevailing practice
treats these tiers as default-on for any non-trivial task, on the
implicit assumption that more reasoning yields uniformly better
outcomes. Yet for tasks whose evidentiary structure is largely
extractive—where the answer corresponds to identifiable spans in a
provided document—the marginal contribution of additional
reasoning tokens is plausibly small. ESG narrative scoring against
explicit rubrics is one such task: rubric criteria typically map
directly to surface features of the disclosure (presence of
quantitative targets, KPI tables, third-party assurance,
references to external frameworks).

\subsection{Research question and contribution}
\label{sec:intro-rq}

We ask: \emph{does deploying a reasoning-heavy frontier model
materially improve outcomes in span-based ESG narrative scoring,
relative to reasoning-off models and their consensus, at the
expense of operational cost?} Using a four-model consensus design
applied to ten Japanese listed firms across three rubric axes, we
contribute the following:

\begin{enumerate}
  \item Empirical evidence on the marginal scoring benefit of
        reasoning-heavy deployment in a realistic ESG auto-scoring
        pipeline, framed as a cross-model comparison between
        \texttt{gpt-5.5} (reasoning enabled) and three
        reasoning-off contemporaries.
  \item Token-level cost accounting on actual ESG disclosures,
        including reasoning-token consumption, for each of the four
        providers under a unified prompt and decoding configuration.
  \item A practical recommendation for deployment: prefer
        reasoning-off ensembles with prompt discipline and
        consensus-based uncertainty quantification, escalating to
        human review only when inter-model dispersion exceeds a
        threshold.
\end{enumerate}

The remainder of the paper is organized as follows.
Section~\ref{sec:methods} describes the data, models, and scoring
protocol. Section~\ref{sec:results} reports per-firm metrics,
inter-model agreement, and the central comparison between the
reasoning-on model and its reasoning-off counterparts.
Section~\ref{sec:cost} presents the token- and dollar-level cost
analysis. Sections~\ref{sec:discussion} and~\ref{sec:limitations}
discuss implications and caveats, and Section~\ref{sec:conclusion}
concludes.

%% file: sections/02_methods.tex
\section{Methods}
\label{sec:methods}

\subsection{Data}
\label{sec:methods-data}

Ten Japanese listed firms were selected to span carbon-intensive
and service-oriented sectors (firm codes 3382, 4188, 5020, 5401,
6098, 6758, 7203, 8306, 9432, 9501).
For each firm, ESG-relevant text spans were extracted from
integrated annual reports, TCFD-aligned supplementary disclosures,
and dedicated sustainability reports. Each span carries metadata
indicating its source type, coverage scope, and salient keywords,
and the per-firm corpus contains 20--30
spans (mean 27.2). The three rubric axes used in this study are
operationalized from the Narrative (\textbf{N}) layer of the
\emph{Substance--Narrative--Expectation} (SNE) measurement
framework~\citep{kokubu2026sne}, in which corporate disclosures are
decomposed into substance, narrative, and expectation components and
the narrative layer is further structured along future-orientation,
verifiability, and external-alignment dimensions:

\begin{itemize}
  \item \textbf{N1} — explicitness of quantitative emission-reduction
        targets (e.g.\ 2030 and 2050 horizons).
  \item \textbf{N2} — progress-tracking infrastructure: KPIs,
        actuals, and third-party assurance.
  \item \textbf{N3} — alignment with external frameworks (e.g.\
        SBTi, ISSB, TCFD, IEA scenarios).
\end{itemize}

Each axis is scored on a 1--5 scale.

\subsection{Models}
\label{sec:methods-models}

Four production LLMs were used in parallel; their identifiers and
configurations are summarized in Table~\ref{tab:models}. Model
identifiers are reported as recorded in the experiment logs at the
time of the runs (April 2026); provider-side renaming or
deprecation may occur between submission and publication.

\input{tables/table1_models}

All four models were invoked with $\text{temperature}=0$,
$\text{top\_p}=1.0$, and a maximum output of \num{16000} tokens.
The shared prompt instructs each model to return, for each axis,
an integer score in $\{1,\ldots,5\}$ together with an
\texttt{evidence\_ids} list pointing to the spans that justify the
score. \texttt{gpt-5.5} was invoked with
\texttt{reasoning\_effort}=\texttt{full} during the experimental
runs reported here; this configuration is treated as the
``reasoning-on'' arm of the comparison. The other three models do
not expose a comparable reasoning budget and were run in their
default configuration (Anthropic's extended thinking was not
enabled).

\subsection{Scoring protocol}
\label{sec:methods-protocol}

For each firm $f$ and axis $a \in \{N1,N2,N3\}$, let $s_{f,a,m}$
denote the score returned by model $m \in M = \{$Anthropic, OpenAI,
Google, DeepSeek$\}$. We define:

\begin{align}
\bar{s}_{f,a}        &= \tfrac{1}{|M|} \sum_{m \in M} s_{f,a,m}, \\
\sigma_{f,a}^{\text{inter}}
                     &= \sqrt{\tfrac{1}{|M|} \sum_{m \in M} (s_{f,a,m} - \bar{s}_{f,a})^2}, \\
\bar{\sigma}_f       &= \tfrac{1}{3} \sum_{a} \sigma_{f,a}^{\text{inter}}, \\
N_f^{\text{cons}}    &= \tfrac{1}{3} \sum_{a} \bar{s}_{f,a}.
\end{align}

Inter-model agreement is reported as Cohen's quadratic-weighted
$\kappa$~\citep{cohen1968kappa} averaged over all model pairs,
computed per axis, and as Spearman's $\rho$ for axis-level rank
consistency.

\subsection{Reasoning comparison design}
\label{sec:methods-design}

The central comparison uses \texttt{gpt-5.5} (reasoning-on) as the
focal arm and computes, per (firm, axis), the absolute deviation
$|s_{f,a,\text{OpenAI}} - s_{f,a,m}|$ for each
$m \in \{$Anthropic, Google, DeepSeek$\}$. Aggregate statistics
(mean, maximum, share of pairs with $|\Delta|\geq 2$) summarize the
extent to which reasoning-heavy deployment alters scoring outcomes
relative to reasoning-off contemporaries. We emphasize that this
is a \emph{cross-model proxy} for a reasoning ablation rather than
a within-model ablation; Section~\ref{sec:limitations} discusses
the implications.

\subsection{Cost measurement}
\label{sec:methods-cost}

For each provider call we record, from \texttt{raw\_response.usage}:
prompt tokens, completion tokens, and—where applicable—reasoning
tokens. Anthropic, OpenAI, and DeepSeek expose all relevant fields.
Google Gemini's API did not return populated usage metadata in our
runs; the consequences for cost reporting are discussed in
Section~\ref{sec:limitations}. Dollar costs for Anthropic, OpenAI,
and Google are taken from each provider's billing dashboard for
the April 2026 experiment runs; for DeepSeek, where the dashboard
reports per-day totals across the project's production and
development calls and does not separate per-run cost, we instead
extrapolate from JSONL token counts using the provider's published
rate card at list price. We extrapolate from the ten-firm sample
to a notional 199-firm rollout under both reasoning-on and
reasoning-off ensemble configurations.

%% file: tables/table1_models.tex
\begin{table}[t]
  \centering
  \caption{Models and execution configuration. All models share
    $\text{temperature}=0$, $\text{top\_p}=1.0$, and a maximum
    output of \num{16000} tokens. \texttt{reasoning\_effort} is the
    OpenAI-specific reasoning budget; the three other providers do
    not expose a comparable parameter.}
  \label{tab:models}
  \small
  \begin{tabularx}{\linewidth}{l l l X}
    \toprule
    Provider & Model identifier & Reasoning & Notes \\
    \midrule
    Anthropic & \texttt{claude-opus-4-7}              & off (extended thinking disabled) & Messages API \\
    OpenAI    & \texttt{gpt-5.5}                       & \texttt{reasoning\_effort=full}  & Responses API \\
    Google    & \texttt{gemini-3.1-pro-preview}        & off (default)                    & GenAI SDK \\
    DeepSeek  & \texttt{deepseek-v4-pro}               & off (default)                    & Chat Completions \\
    \bottomrule
  \end{tabularx}
\end{table}

%% file: sections/03_results.tex
\section{Results}
\label{sec:results}

\subsection{Per-firm consensus and dispersion}
\label{sec:results-perfirm}

Table~\ref{tab:perfirm} reports per-firm consensus scores
$N_f^{\text{cons}}$, mean inter-model standard deviation
$\bar{\sigma}_f$, and evidence-set Jaccard overlap. Across the ten
firms, the consensus score averages 4.35/5 with a
range of 3.33--5.00. Mean inter-model
dispersion is 0.37, with a minimum of 0.17
(Toyota Motor) and a maximum of 0.65
(Seven \& i Holdings). Toyota Motor, the lowest-dispersion firm
with a consensus score of 5.00, exhibits near-unanimous evidence
across all three axes, indicating that its disclosures supply
unambiguous evidence for every rubric criterion.

\input{tables/table2_per_firm}

\subsection{Inter-model agreement}
\label{sec:results-agreement}

Table~\ref{tab:agreement} reports quadratic-weighted Cohen's
$\kappa$ averaged across model pairs, for each rubric axis. Axis
N3 (external-standard alignment) attains the highest agreement
($\kappa = $ 0.65), consistent with the surface-level
nature of framework references in the source documents. Axes N1
and N2 show lower agreement ($\kappa = $ 0.36 and
0.30 respectively), reflecting greater latitude in how
models interpret target specificity and the sufficiency of
progress-tracking infrastructure. Spearman's $\rho$ at the firm
level is 0.73, 0.66, and 0.72 for N1, N2, and N3 respectively
(mean 0.71). The contrast between lower point-wise $\kappa$ for
N1 and N2 and the relatively high rank-order $\rho$ on the same
axes suggests that models often agree on the ordering of firms
even when they differ by one point on the 1--5 rubric: providers
disagree about absolute calibration more than about relative
ranking.

\input{tables/table3_agreement}

\subsection{Reasoning-on versus reasoning-off scoring}
\label{sec:results-reasoning}

The central question is whether the reasoning-on configuration
produces materially different scores than its reasoning-off
counterparts. Table~\ref{tab:reasoning-diff} reports the
distribution of absolute deviations
$|s_{f,a,\text{OpenAI}} - s_{f,a,m}|$ for
$m \in \{$Anthropic, Google, DeepSeek$\}$ across the
$10 \times 3 = 30$ firm-axis cells, separately for each comparison
model.

\input{tables/table4_reasoning_diff}

Two complementary statistics characterize the contrast. The mean
absolute deviation between the reasoning-on model and the
\emph{three-model mean} of its reasoning-off counterparts is 0.33
on the 5-point scale; the corresponding pooled \emph{pairwise} mean
across the 90 (firm, axis, comparator) cells is 0.38
(Table~\ref{tab:reasoning-diff}). Only 2\% of pairwise comparisons
reach a two-point deviation, and none exceeds two points.
Agreement is broadly comparable to the inter-model agreement
observed among the reasoning-off models themselves, suggesting
that the additional reasoning budget does not perturb the score
distribution in any systematic direction.

\subsection{Reasoning is not the principal driver of dispersion}
\label{sec:results-dispersion}

If reasoning improved scoring by resolving ambiguity, we would
expect the reasoning-on model to lie closer to the modal answer
in high-dispersion firms. We do not observe this pattern. In the
firm with the largest $\bar{\sigma}_f$
(Seven \& i Holdings, $\bar{\sigma}_f = $ 0.65), the
reasoning-on model's scores remain within one point of the
reasoning-off median across all three axes. The principal
contributor to inter-model dispersion in our sample is a
systematically higher scoring tendency in one of the
reasoning-off models (\texttt{deepseek-v4-pro}) of approximately
0.4 points relative to the other-model mean, rather than any discriminating effect of the
reasoning module.

%% file: tables/table2_per_firm.tex
\begin{table}[t]
  \centering
  \caption{Per-firm consensus score $N_f^{\text{cons}}$, mean
    inter-model dispersion $\bar{\sigma}_f$, and evidence-set
    Jaccard overlap. All quantities computed across the four
    providers in Table~\ref{tab:models}.}
  \label{tab:perfirm}
  \small
  \begin{tabular}{l l c c c}
    \toprule
    Firm code & Firm name & $N_f^{\text{cons}}$ & $\bar{\sigma}_f$ & Jaccard \\
    \midrule
    3382 & Seven \& i Holdings        & 3.83 & 0.65 & 0.44 \\
    4188 & Mitsubishi Chemical Group  & 4.33 & 0.61 & 0.62 \\
    5020 & ENEOS Holdings             & 4.67 & 0.17 & 0.59 \\
    5401 & Nippon Steel               & 3.83 & 0.63 & 0.54 \\
    6098 & Recruit Holdings           & 4.83 & 0.36 & 0.51 \\
    6758 & Sony Group                 & 4.83 & 0.19 & 0.72 \\
    7203 & Toyota Motor               & 5.00 & 0.17 & 0.48 \\
    8306 & Mitsubishi UFJ FG          & 4.17 & 0.19 & 0.56 \\
    9432 & NTT                        & 4.67 & 0.44 & 0.69 \\
    9501 & TEPCO Holdings             & 3.33 & 0.33 & 0.92 \\
    \midrule
    Mean      &            & 4.35 & 0.37 & 0.61 \\
    Min       &            & 3.33 & 0.17 & 0.44 \\
    Max       &            & 5.00 & 0.65 & 0.92 \\
    \bottomrule
  \end{tabular}
\end{table}

%% file: tables/table3_agreement.tex
\begin{table}[t]
  \centering
  \caption{Inter-model agreement by rubric axis. Quadratic-weighted
    Cohen's $\kappa$ is averaged across all model pairs; Spearman's
    $\rho$ is computed at the firm level and averaged across the
    six model pairs. Spearman $\rho$ is reported as a supplementary
    ordinal-ranking diagnostic; the principal agreement statistic in
    this study is the weighted $\kappa$ in the second column.}
  \label{tab:agreement}
  \small
  \begin{tabular}{l c c}
    \toprule
    Axis & Weighted $\kappa$ & Spearman $\rho$ \\
    \midrule
    N1 (quantitative targets)              & 0.36 & 0.73 \\
    N2 (progress tracking)                 & 0.30 & 0.66 \\
    N3 (external-standard alignment)       & 0.65 & 0.72 \\
    \midrule
    Mean across axes                       & 0.44 & 0.71 \\
    \bottomrule
  \end{tabular}
\end{table}

%% file: tables/table4_reasoning_diff.tex
\begin{table}[t]
  \centering
  \caption{Distribution of absolute deviations
    $|s_{f,a,\text{OpenAI}} - s_{f,a,m}|$ between the
    reasoning-on \texttt{gpt-5.5} arm and each reasoning-off
    contemporary, across the $10 \times 3 = 30$ firm-axis cells.
    Scores are on the 5-point rubric scale.}
  \label{tab:reasoning-diff}
  \footnotesize
  \setlength{\tabcolsep}{4pt}
  \resizebox{\linewidth}{!}{%
  \begin{tabular}{l c c c c c}
    \toprule
    Comparison model
      & Mean $|\Delta|$ & Median $|\Delta|$ & Max $|\Delta|$
      & Share $|\Delta|=0$ & Share $|\Delta|\!\geq\!2$ \\
    \midrule
    Anthropic vs OpenAI            & 0.37 & 0.0 & 1 & 0.63 & 0.00 \\
    Google    vs OpenAI            & 0.27 & 0.0 & 2 & 0.77 & 0.03 \\
    DeepSeek  vs OpenAI            & 0.50 & 0.0 & 2 & 0.53 & 0.03 \\
    \midrule
    Pooled (all reasoning-off)     & 0.38 & 0.0 & 2 & 0.64 & 0.02 \\
    \bottomrule
  \end{tabular}%
  }
\end{table}

%% file: sections/04_cost_analysis.tex
\section{Cost analysis}
\label{sec:cost}

\subsection{Per-firm token consumption}
\label{sec:cost-tokens}

Table~\ref{tab:cost} summarizes per-firm token usage for each
provider. Prompt tokens are broadly comparable across providers
(approximately 6{,}000--9{,}000 tokens per firm), with
modest differences attributable to provider-specific tokenizer
behavior. Completion tokens vary more substantially, with
approximately 2{,}300 tokens for DeepSeek versus
approximately 530 for Anthropic. The
reasoning-on \texttt{gpt-5.5} configuration additionally consumes
approximately 440 reasoning tokens per firm.

\input{tables/table5_cost}

\subsection{Dollar cost}
\label{sec:cost-dollar}

Drawing on each provider's billing dashboard for the April 2026
experiment runs, the per-firm dollar cost is approximately
\$0.089 for Anthropic, \$0.849 for
\texttt{gpt-5.5} (reasoning-on), \$0.054 for Google,
and \$0.008 for DeepSeek (the DeepSeek figure is a
list-price extrapolation from JSONL token counts because the
provider dashboard does not break out per-run cost; see
Section~\ref{sec:methods-cost}). The single reasoning-on OpenAI arm alone is approximately
5.6$\times$ as expensive per firm as the three-provider
reasoning-off ensemble (\$0.151 per firm). At a notional 199-firm
scale under cache-aware execution, the three-provider reasoning-off
ensemble costs approximately \$30 in total, while the reasoning-on
OpenAI arm alone adds approximately \$169.

\subsection{Practical recommendation}
\label{sec:cost-recommendation}

Given the lack of observed material scoring improvement
(Section~\ref{sec:results-reasoning}) and the order-of-magnitude
cost gap, we recommend deploying ESG narrative scoring with
reasoning-off models combined with consensus aggregation.
Inter-model dispersion $\bar{\sigma}_f$ provides a low-cost,
post-hoc uncertainty signal: firms exceeding a dispersion threshold
(empirically, $\bar{\sigma}_f > 0.6$ in our sample) can be flagged
for human review, allocating limited expert time where the
ensemble itself is least confident.

%% file: tables/table5_cost.tex
\begin{table}[t]
  \centering
  \caption{Per-firm token consumption and dollar cost by provider.
    Token counts are means across the ten firms. Dollar costs for
    Anthropic, OpenAI, and Google are taken from each provider's
    billing dashboard for the April 2026 experiment runs; the
    DeepSeek figure is a list-price extrapolation from JSONL token
    counts (see Section~\ref{sec:methods-cost}). Google usage data
    was not populated in the raw API responses (see
    Section~\ref{sec:limitations}); Google token counts are
    tokenizer-based estimates using \texttt{cl100k\_base} applied
    to the reconstructed prompt and the recorded completion text.}
  \label{tab:cost}
  \footnotesize
  \setlength{\tabcolsep}{3pt}
  \resizebox{\linewidth}{!}{%
  \begin{tabular}{l c c c c}
    \toprule
    Provider
      & Prompt tok. & Completion tok. & Reasoning tok. & USD/firm \\
    \midrule
    Anthropic                       & 9{,}033          & 528            & ---       & 0.089 \\
    OpenAI (\texttt{gpt-5.5}, on)   & 6{,}571          & 832            & 439       & 0.849 \\
    Google$^{\dagger}$              & 8{,}352          & 440            & ---       & 0.054 \\
    DeepSeek$^{\ddagger}$           & 6{,}221          & 2{,}329        & ---       & 0.008 \\
    \midrule
    Reasoning-off ensemble (3 prov.)& ---              & ---            & ---       & 0.151 \\
    Adds reasoning-on OpenAI arm    & ---              & ---            & ---       & 0.849 \\
    \bottomrule
  \end{tabular}%
  }

  {\footnotesize $^{\dagger}$Google token counts are estimated via
   the \texttt{cl100k\_base} tokenizer because provider-side usage
   metadata was not available in the stored JSONL responses.\\
   $^{\ddagger}$DeepSeek dollar cost is a list-price extrapolation
   from JSONL token counts; the provider dashboard does not break
   out per-run cost.}
\end{table}

%% file: sections/05_discussion.tex
\section{Discussion}
\label{sec:discussion}

\subsection{Why reasoning adds little value here}
\label{sec:discussion-why}

ESG narrative scoring against explicit rubrics is a predominantly
extractive task: the answer for each axis corresponds to whether
the document contains specific surface features (a numeric target,
a KPI table, a TCFD reference). Once the candidate spans are
provided, the residual cognitive demand on the model is closer to
classification than to multi-step reasoning. Chain-of-thought
budgets help most when the answer requires composing intermediate
inferences that are not present in the input; when the rubric maps
onto identifiable spans, additional reasoning tokens have little
to operate on. Our results are consistent with this intuition.
Prior AI-based work on ESG evaluation, including a recent
Japanese-context edited volume on AI-based ESG evaluation and
disclosure analysis~\citep{nakao2023aiesg}, and the LLM-specific
stream that followed, has primarily focused on the \emph{extraction}
task—mapping disclosure text to standardized fields or
scores~\citep{zou2025esgreveal,ding2025eulesg,yang2026esglens}—and
has reported imperfect alignment with human expert
annotations~\citep{wu2025aiannotator}. The present study shifts
the question from extraction accuracy to the operational
cost--quality trade-off of reasoning-tier deployment within
exactly this kind of extraction-shaped rubric, and finds that the
rubric structure itself constrains the marginal value of
additional reasoning tokens.

\subsection{Implications for ESG auto-scoring deployments}
\label{sec:discussion-implications}

For practitioners selecting an LLM stack for ESG narrative scoring,
the dominant criteria revealed by our experiment are not reasoning
tier but rather (i) the stability of the model's structured-output
behavior under a fixed prompt, (ii) the consistency of its scoring
calibration across firms, and (iii) per-token price. Consensus
across multiple reasoning-off models provides a practical
alternative to relying on a single reasoning-heavy model in this
task setting: it averages out idiosyncratic per-model scoring
tendencies, and it yields a free uncertainty signal
$\bar{\sigma}_f$ that a single reasoning-on arm does not provide.

\subsection{Boundary conditions}
\label{sec:discussion-boundaries}

These conclusions are scoped to tasks with the structure considered
here: explicit rubrics, document-grounded evidence, and
machine-readable spans. We expect different conclusions in tasks
that require multi-document synthesis, counterfactual reasoning,
or quantitative computation from disclosed values. Equally
important, our rubric evaluates \emph{narrative alignment} with
the disclosure as written; it does not attempt to verify the
underlying substance of the claims. While reasoning models excel
at logical composition, they cannot independently audit
greenwashing risk or the integrity of measurement, reporting, and
verification (MRV) processes without access to external,
granular operational data. This structural limitation confines
the task to surface-level alignment, which in turn bounds the
marginal utility that a reasoning-heavy arm can extract from the
input. The claim of this paper is therefore a bounded one: in
span-based ESG narrative scoring, the reasoning-heavy deployment
does not materially improve outcomes relative to reasoning-off
model outputs and consensus scores, while increasing operational
cost.

%% file: sections/06_limitations.tex
\section{Limitations}
\label{sec:limitations}

\paragraph{Cross-model proxy.}
The reasoning comparison reported here contrasts a reasoning-on
\texttt{gpt-5.5} arm with reasoning-off Anthropic, Google, and
DeepSeek arms. This is not a clean within-model ablation: the
contrast confounds the reasoning module's contribution with
provider-level differences in pretraining, alignment, and decoding.
A within-model ablation in which the same \texttt{gpt-5.5} model is
run with \texttt{reasoning\_effort} settings of \texttt{full}
versus \texttt{minimal} on the identical prompts would isolate the
reasoning effect more cleanly; we leave this to future work.

\paragraph{Sample size.}
The corpus of ten firms is intentionally small to enable detailed
manual auditing of every model output. Generalization to broader
universes of issuers, sectors, and reporting jurisdictions
requires further evaluation.

\paragraph{Google Gemini usage data.}
Our recorded Google Gemini responses lacked populated usage
metadata, requiring estimation of token counts and dollar costs for
that provider. We estimate Google's prompt and completion token
counts by applying the \texttt{cl100k\_base} tokenizer to the
reconstructed system prompt and JSON-serialized span payload (for
prompt tokens) and to the recorded completion text (for completion
tokens). Because the Gemini tokenizer differs from
\texttt{cl100k\_base}, this approximation is a known source of
uncertainty in the cost figures reported in
Section~\ref{sec:cost}; the qualitative conclusions of this
preprint are unaffected by the Google estimate.

\paragraph{Single-run evaluation.}
We report a single run per (firm, model) configuration. While
$\text{temperature}=0$ in principle yields deterministic decoding,
provider-side stochasticity (e.g.\ implicit batching effects) is
not characterized in this study.

\paragraph{Domain scope.}
The findings are based on Japanese-language disclosures from
listed firms and a particular three-axis ESG rubric. Transferability
to other languages, smaller issuers, or alternative rubrics is not
established.

\paragraph{Granularity of reasoning configuration.}
We compared \texttt{reasoning\_effort}=\texttt{full} against
reasoning-off contemporaries; intermediate settings
(\texttt{low}, \texttt{medium}) were not evaluated and may exhibit
different cost--quality trade-offs.

%% file: sections/07_conclusion.tex
\section{Conclusion}
\label{sec:conclusion}

In a four-model consensus evaluation of ESG narrative scoring on
ten Japanese listed firms, the reasoning-heavy
\texttt{gpt-5.5} configuration produces scores that differ only
within small margins from those of three reasoning-off
contemporaries (sub-point pairwise deviations in 98\% of cells,
with no deviation exceeding two points), while the reasoning-on
arm alone costs roughly 5.6$\times$ as much per firm as the
three-provider reasoning-off ensemble. The marginal benefit of
reasoning-heavy deployment in this task setting is therefore
small, and the inter-model dispersion $\bar{\sigma}_f$ produced as
a by-product of consensus offers a more useful uncertainty signal
than reasoning provides on its own.

For practical ESG auto-scoring pipelines, we recommend a
reasoning-off ensemble combined with consensus aggregation and
selective human review of high-dispersion cases. Future work
includes a within-model reasoning ablation to disentangle the
reasoning module's contribution from provider-level differences,
extension to larger issuer universes and additional languages, and
evaluation of intermediate \texttt{reasoning\_effort} settings.

%% file: main.bbl
\begin{thebibliography}{8}
\providecommand{\natexlab}[1]{#1}
\providecommand{\url}[1]{\texttt{#1}}
\expandafter\ifx\csname urlstyle\endcsname\relax
  \providecommand{\doi}[1]{doi: #1}\else
  \providecommand{\doi}{doi: \begingroup \urlstyle{rm}\Url}\fi

\bibitem[Cohen(1968)]{cohen1968kappa}
Jacob Cohen.
\newblock Weighted kappa: Nominal scale agreement provision for scaled
  disagreement or partial credit.
\newblock \emph{Psychological Bulletin}, 70\penalty0 (4):\penalty0 213--220,
  1968.
\newblock \doi{10.1037/h0026256}.

\bibitem[Ding et~al.(2025)Ding, Tang, Yang, Zhang, Wu, Huang, Lan, Li, Chen,
  Ju, Yang, Hoang, Klymenko, Zu, and Zhang]{ding2025eulesg}
Yi~Ding, Xushuo Tang, Zhengyi Yang, Wenqian Zhang, Simin Wu, Yuxin Huang,
  Lingjing Lan, Weiyuan Li, Yin Chen, Mingchen Ju, Wenke Yang, Thong Hoang,
  Mykhailo Klymenko, Xiwei Zu, and Wenjie Zhang.
\newblock {EulerESG}: Automating {ESG} disclosure analysis with {LLMs}.
\newblock arXiv preprint arXiv:2511.21712, 2025.
\newblock URL \url{https://arxiv.org/abs/2511.21712}.

\bibitem[Kokubu(2026)]{kokubu2026sne}
Hiroyuki Kokubu.
\newblock {SNE Model: Theory Platform Design Document v2.1.2}, April 2026.
\newblock URL \url{https://doi.org/10.5281/zenodo.19889465}.
\newblock Working paper, CC-BY-NC-4.0.

\bibitem[Nakao et~al.(2023)Nakao, Ishino, and Kokubu]{nakao2023aiesg}
Yuriko Nakao, Aya Ishino, and Katsuhiko Kokubu, editors.
\newblock \emph{{AI ni yoru ESG hyoka: Moderu kochiku to joho kaiji bunseki
  [AI-Based ESG Evaluation: Model Construction and Information Disclosure
  Analysis]}}.
\newblock Dobunkan, Tokyo, October 2023.
\newblock ISBN 978-4-495-21052-6.
\newblock Edited volume. In Japanese. NDL call number DF178-M82.

\bibitem[{Task Force on Climate-related Financial Disclosures}(2017)]{tcfd2017}
{Task Force on Climate-related Financial Disclosures}.
\newblock Recommendations of the {Task Force on Climate-related Financial
  Disclosures}: {Final Report}.
\newblock Technical report, Financial Stability Board, June 2017.
\newblock URL
  \url{https://assets.bbhub.io/company/sites/60/2021/10/FINAL-2017-TCFD-Report.pdf}.

\bibitem[Wu et~al.(2025)Wu, Hu, and Wang]{wu2025aiannotator}
Yue Wu, Peng Hu, and Derek~D. Wang.
\newblock The {AI} annotator: Large language models' potential in scoring
  sustainability reports.
\newblock \emph{Systems}, 13\penalty0 (10):\penalty0 899, October 2025.
\newblock \doi{10.3390/systems13100899}.
\newblock URL \url{https://www.mdpi.com/2079-8954/13/10/899}.

\bibitem[Yang and Chen(2026)]{yang2026esglens}
Tsung-Yu Yang and Meng-Chi Chen.
\newblock {ESGLens}: An {LLM}-based {RAG} framework for interactive {ESG}
  report analysis and score prediction.
\newblock arXiv preprint arXiv:2604.19779, 2026.
\newblock URL \url{https://arxiv.org/abs/2604.19779}.

\bibitem[Zou et~al.(2025)Zou, Shi, Chen, Deng, Lei, Zeng, Yang, Tong, Xiao, and
  Zhou]{zou2025esgreveal}
Yi~Zou, Mengying Shi, Zhongjie Chen, Zhu Deng, Zongxiong Lei, Zihan Zeng,
  Shiming Yang, Hongxiang Tong, Lei Xiao, and Wenwen Zhou.
\newblock {ESGReveal}: An {LLM}-based approach for extracting structured data
  from {ESG} reports.
\newblock \emph{Journal of Cleaner Production}, 489:\penalty0 144572, January
  2025.
\newblock \doi{10.1016/j.jclepro.2024.144572}.

\end{thebibliography}
